\documentclass[a4paper,12pt]{article}
\usepackage{epsfig}
\usepackage{citesort}
\date{\today}
 \normalsize

\newcommand{\bmat}{\left(\begin{array}}
\newcommand{\emat}{\end{array}\right)}
\newcommand{\be}{\begin{equation}}
\newcommand{\ee}{\end{equation}}
\newcommand{\bea}{\begin{eqnarray}}
\newcommand{\eea}{\end{eqnarray}}

\def\lsim{\raise0.3ex\hbox{$\;<$\kern-0.75em\raise-1.1ex\hbox{$\sim\;$}}}
\def\gsim{\raise0.3ex\hbox{$\;>$\kern-0.75em\raise-1.1ex\hbox{$\sim\;$}}}

\def\dofigs#1#2#3{\centerline{\epsfxsize=#1\epsfig{file=#2, width=7.5cm, 
height=7.5cm, angle=-90}
\hfil\epsfxsize=#1\epsfig{file=#3,  width=7.5cm, height=7.5cm, angle=-90}}}
\newcommand{\tgb}{\tan{\beta}}

\begin{document}
\renewcommand{\thefootnote}{\fnsymbol{footnote}}
\rightline{IPPP/01/64} \rightline{DCPT/01/128}
\rightline{HIP-2001-67/TH} \rightline{CERN-TH/2002-001}
\vspace{.3cm} 
{\Large
\begin{center}
{\bf On the $B \to X_s\,l^+ l^-$ decays in general\\ supersymmetric models }
\end{center}}
\vspace{.3cm}

\begin{center}
E. Gabrielli$^{1,2}$ and S. Khalil$^{3,4}$\\
\vspace{.3cm}
$^1$\emph{ Theory Division, CERN, CH-1211 Geneva 23, Switzerland}
\\
$^2$\emph{
Helsinki Institute of Physics,
     POB 64,00014 University of Helsinki, Finland}
\\
$^3$ \emph{IPPP, Physics Department, Durham University, DH1 3LE,
Durham,~~U.~K.}
\\
$^4$ \emph{Ain Shams University, Faculty of Science, Cairo, 11566,
Egypt.}

\end{center}

\vspace{.3cm}
\hrule \vskip 0.3cm
\begin{center}
\small{\bf Abstract}\\[3mm]
\end{center}
We analyze the inclusive semileptonic decays $B \to X_s\, l^+ l^-$ 
in the framework of the supersymmetric standard model
with non-universal soft-breaking terms at GUT scale.
We show that the general trend of universal and 
non-universal models is a decreasing of branching ratio (BR) 
and increasing of energy asymmetry (AS).
However, only non--universal models can have chances to get very
large enhancements in BR and AS, corresponding to large
(negative) SUSY contributions to the $b \to s \gamma$ amplitude.

\begin{minipage}[h]{14.0cm}
\end{minipage}
\vskip 0.3cm \hrule \vskip 0.5cm

Flavor changing neutral current (FCNC) and CP violating phenomena
can be considered as one of the best indirect probe for physics 
beyond the standard model (SM).
Due to the absence of tree level FCNCs in the SM and the 
suppression of the Glashow-Iliopoulos-Maiani mechanism,
they are particularly sensitive to any non standard physics contribution.

In the framework of low energy supersymmetric (SUSY) models, FCNC processes
play an important role in severely constraining the soft breaking
sector of supersymmetry \cite{FCNCsusy}.
As known, these constraints require an high degree of 
degeneracy in the squark mass matrices, suggesting that 
the mechanism which transmits the SUSY breaking 
to the observable sector should be flavour blind.
For instance, minimal supergravity (mSUGRA) and gauge--mediation mechanisms 
successfully explain this degeneracy.
In particular, in mSUGRA scenario 
all the tests on FCNCs can be satisfied due
to the assumption of universality for the soft breaking terms at GUT scale.
However, recently there has been a growing interest concerning 
supersymmetric models with non--universal SUSY soft--breaking terms
\cite{Brignole:1997fb}.
This is motivated by the fact that superstring inspired models,
where supergravity theories are derived,
naturally favour non--universality in the soft-breaking terms \cite{Ibanez}.
This is mainly due 
to the fact that superstring theories live in extra--dimensions and 
after compactification, non--flat K\"ahler metrics
and flavour--dependent SUSY soft--breaking terms can arise.

Particularly interesting among this class of models are the ones
with non--universal trilinear soft--breaking terms in the scalar sector, 
the so--called A--terms. These models can have interesting phenomenological
consequences. They could solve in principle 
the SUSY CP problem, satisfy all the FCNC 
constraints, and provide at the same time 
new significant contributions to the direct CP violating parameter 
$\varepsilon^{\prime}/\varepsilon$ as suggested by 
the recent experimental results on $\varepsilon^{\prime}/\varepsilon$
~\cite{nonuniversalA}.
Moreover, it has been argued that this class of 
models should also pass the strong constraint on $B \to X_s\, \gamma$ 
decay~\cite{Emidio} and gives rise to a large contribution to the CP 
asymmetry, of order $10\% - 15\%$ which can be accessible at $B$ factories
~\cite{bailin}.

In this letter we analyze the impact of a large class of 
supersymmetric models with non--universal soft SUSY breaking terms 
(which is motivated by the string inspired scenarios)
in the semileptonic (inclusive) $B\to X_s\, l^+l^-$ decays (with 
$l=e, \mu$). As for the $B \to X_s \gamma$ decay, these 
processes are also very interesting for several reasons: first they are 
very sensitive to large $\tgb$ since they involve the
magnetic dipole operator $Q_7$ (see Eq.(\ref{operator7})) 
which allows the quark 
$b\to s \gamma$ transition.
Second, they involve other operators as well, the semileptonic
operators $Q_9$ and $Q_{10}$ (see Eqs.(\ref{operator9}-\ref{operator10})), 
and so can serve 
as complementary tests of the model. Third, they provide several measurable
quantities, such as branching ratios and asymmetries. At present 
these decays are known in QCD at the next--to--leading (NLO) order logaritmhmic
accuracy for the SM \cite{bsllSM}, and also $1/m_b$ 
nonperturbative contributions are small and well under control.

From the experimental side, the situation about these decay channels 
is quite exciting.
The BELLE experiment has recently announced the first evidence 
for the exclusive process $B\to K^{\star} l^+l^-$~\cite{belle}, 
and upper bounds 
for the three body decays $B\to (K, K^{\star}) + (e^+e^-, \mu+\mu-)$, 
reported by BABAR and BELLE, are very close the SM expectations
~\cite{babar,belle}. However, exclusive processes are affected 
by larger theoretical uncertainties than the inclusive ones due to 
model dependent calculations of hadronic matrix elements.
For this reason we will restrict our analysis to the inclusive ones.

In the framework of supersymmetric models, there are several studies 
about $B\to X_s\, l^+l^-$ decays in the literature
~\cite{masiero,ali,goto,BBMR,misiak,sarid}.
However, a detailed analysis about SUSY models with non--universal
soft breaking terms at GUT scale has not been considered.
As suggested by a recent study~\cite{masiero}, 
based on the low energy approach to
supersymmetric models, the non--universality in the soft--breaking sector
could generate significant departures from the SM in 
the semileptonic $B\to X_s\, l^+l^-$ decays.
In this analysis the mass insertion method has been used, where the
pattern of flavour change is parametrized by the ratios
\be
(\delta_{i j})^f_{AB}= \frac{(m^2_{ij})^f_{AB}}{M_{sq}^2}
\ee
where $(m^2_{ij})^f_{AB}$ are the off--diagonal elements of the
the $f=\tilde{u},\tilde{d}$ scalar mass squared matrix which mixes flavour 
$i,j$ for both left-- and right--handed scalars ($A,B=\mathrm{left}, 
\mathrm{right})$, and $M_{sq}$ is the average squark mass.
The main conclusion of this work is that FCNCs constraints 
and vacuum stability bounds, which strongly constrain these $\delta$s, 
could not prevent large effects in $B\to X_s\, l^+l^-$ decays. 
In particular, large SUSY contributions to  
the Wilson coefficients $C_9$ and $C_{10}$ at EW scale, corresponding 
respectively to the semileptonic operators
$Q_9$ and $Q_{10}$, are possible. 
Therefore, generic SUSY models (with non--universalities in the scalar 
sector and $A$-terms implemented at GUT scale)
seem indeed an ideal scenario where these large effects could be found.
However, it should be stressed that in the analysis of Ref.\cite{masiero}, 
the enhancement of $C_9$ and $C_{10}$ is obtained by 
taking all the $\delta$s and other 
SUSY parameters at low energy as free parameters,
in particular the gluino, the lightest stop mass and the bilinear Higgs
couplings (the $\mu$ term).
In the class of models analyzed here, we will see that these
sizable effects to $C_9$ and $C_{10}$ will not show up,
leaving to potential large deviations only in the Wilson coefficient ($C_7$)
of the magnetic--dipole operator $Q_7$.
The main reason is due to the fact that 
the relevant (low energy) SUSY parameters for enhancing $C_9$ and $C_{10}$
are strongly correlated, leaving the $B\to X_s \gamma$ and
the experimental bounds on SUSY mass spectrum very effective in 
preventing such enhancements.

Furthermore, we will consider the effect of the SUSY models with 
non--abelian flavour symmetry on these semileptonic decays.
The main effect of this symmetry is to prevent excessive 
FCNC effects in case that the mechanism of SUSY breaking should not be
flavour blind. 
As a specific example, we will analyze here the model proposed in 
Ref.\cite{silvestrini},
in which the pattern of flavour violation is implemented 
by the breaking of an U(2) (horizontal) flavour symmetry.
For the same reason given above, also these models have large potentialities 
to give sizable deviations in $B\to X_s\, l^+l^-$ decays, since they contain 
a new flavour structure in addition to Yukawa matrices.
However, we will see that the same conclusions about 
sizable contributions to $C_9$ and $C_{10}$ will hold for these models
as well.

Now we start with the SM results for the inclusive $B\to X_s\, l^+l^-$ decays.
Inclusive hadronic rates in B meson decays can be precisely 
calculated by using perturbative QCD and $1/m_b$ quark expansion.
The effective Hamiltonian for the $b$ quark semileptonic decay 
$b \to s\, l^+ l^-$ is given by
\be
H_{eff}=-\frac{4G_F}{\sqrt{2}}V_{ts}^{\star}V_{tb}\sum_{i=1}^{10}
C_i(\mu_b) Q_i(\mu_b) \;,
\label{Heff} 
\ee   
where $Q_i(\mu)$ are the $\Delta B=1$ transition operators evaluated at the 
renomalization scale $\mu \simeq \mathcal{O}(m_b)$. 
A complete list of operators involved in this decay are given in 
Refs.\cite{operators,bsllSM}. The relevant operators that can 
be affected by the SUSY contributions are given by
\bea 
Q_7 &=& \frac{e}{16 \pi^2} m_b \bar{s}_L \sigma^{\mu \nu} b_R~ F_{\mu \nu}\;,
\label{operator7}
\\
Q_8 &=& \frac{g_s}{16 \pi^2} m_b \bar{s}_L T^a \sigma^{\mu \nu} b_R~ G_{\mu \nu}\;,
\label{operator8}
\\
Q_9 &=& (\bar{s}_L \gamma_{\mu} b_L)~ \bar{l} \gamma^{\mu} l\;,
\label{operator9}
\\
Q_{10} &=& (\bar{s}_L \gamma_{\mu} b_L)~ \bar{l} \gamma^{\mu} \gamma_5 l\;.
\label{operator10}
\eea
At one-loop, the SUSY contributions to these operators are given by 
$Z$ and $\gamma$ super-penguin and box diagrams, where inside the loop 
can run charged Higgs, charginos, gluinos, neutralinos, squarks, and sleptons 
\cite{BBMR,misiak}.

The general SUSY Hamiltonian also contains the operators $\tilde{Q}_i$ which 
have opposite chirality with respect to the $Q_i$ ones. 
In the SM and minimal flavor SUSY
models, these contributions are suppressed by $\mathcal{O}(m_s/m_b)$. However,
in generic SUSY models, and in particular, in case of non--degenerate 
$A$--terms this argument is no longer true. For instance, the gluino 
contribution to these operators depend on $(\delta^d_{23})_{LR}$ and 
$(\delta^d_{23})_{RL}$. Here both of these mass insertions are linear 
combinations of the down type quark masses rather than $m_b$ or $m_s$ 
exclusively. Therefore, to be consistent, one has to include the 
contributions of these operators. Indeed, the effects of 
the operators $\tilde{Q}_{7,8}$ have been found to be very significant 
for the branching ratio of the $B \to X_s \gamma$ decay~\cite{Emidio} and 
for the CP asymmetry of this decay as well~\cite{bailin}. 
The Wilson coefficients $C_i(\mu)$ can be decomposed as
\be
C_i (\mu) = C_i^{(0)} (\mu) + \frac{\alpha_s(\mu)}{4 \pi} C_i^{(1)}(\mu) +
\mathcal{O}(\alpha_s^2)\;,
\ee
where $C_i^{(0)}$ and $C_i^{(1)}$ refer to the LO
and NLO results, 
respectively. For our purpose, the SUSY corrections from including the NLO and
NNLO are unimportant.  The new physics effects in  $b \to s\, l^+ l^- $ 
can be paramterized by $R_i$ and $\tilde{R}_i$, $i=7,8,9,10$ 
defined at the EW as
\bea
R_i = \frac{C_i^{(0)} -C_i^{(0)SM}}{C_i^{(0)SM}}\;, ~~~~~~~~~~~~~~~~~~~~
\tilde{R}_i = \frac{\tilde{C}_i^{(0)}}{C_i^{(0)SM}}\;.
\eea
Note that there is no SM contribution to $\tilde{C}_i$. 
In the minimal supersymmetric standard model (MSSM), the expressions
for $R_i$ and $\tilde{R}_i$ are given in Refs.\cite{BBMR,misiak,Emidio}.
However, we anticipate that in the class of models analyzed here,
the SUSY contribution to $R_9$ and $R_{10}$ is very small 
in comparison to $R_7$, the same conclusion hold for $\tilde{R}_9$
and $\tilde{R}_{10}$ as well. Therefore, in order to simplify our analysis, 
we will use the approximation in which 
the SUSY dependence in $b\to s\; l^+l^-$ decay enters only
through the ratios of Wilson coefficients
$R_7$, $R_9$, $R_{10}$, and $\tilde{R}_7$. Note that the dependence on 
$R_8$ and $\tilde{R}_8$ is modest in $b\to s\; l^+l^-$, 
due to the fact that the operator $Q_8$ mixes with $Q_7$ at the NLO.
For this reason we will neglect their contribution. We have explicitly 
checked that this approximation does not significantly affect our results.

Using this parametrization,
the non--resonant branching ratios (BR)\footnote{
In order to reduce the large non--perturbative contributions to the 
$B\to X_s\, l^+l^-$ decays, the resonant regions in the final invariant mass
of the  dilepton system $l^+l^-$ should be avoided. This can be easily 
implemented by excluding some special areas from the integration regions 
in the dilepton invariant mass. The resulting
BR where these regions have not been included, is usually 
called the non--resonant BR.}
are expressed in terms of the new physics contribution
as \cite{misiak,sarid}.
\bea
\mathrm{BR}\!\!\!\!&(&\!\!\!\!B\to X_s\, e^+ e^-) = 7.29 \times 10^{-6} 
\Big(1 + 0.714~ R_{10}+ 0.357~ R_{10}^2 + 0.35~ R_7 
\nonumber\\
&+&0.0947(R_7^2+\tilde{R}_7^2) + 0.179 R_9 -0.0313~ R_7 R_9 + 
0.045~ R_9^2 \Big),
\label{BRe}\\
\mathrm{BR}\!\!\!\!&(&\!\!\!\!B\to X_s\, \mu^+ \mu^-) = 4.89 \times 10^{-6}
\Big(1 + 1.07~ R_{10}+ 0.535~ R_{10}^2 + 0.0982~ R_7 
\nonumber\\
&+&0.0491(R_7^2+\tilde{R}_7^2) + 0.264 R_9 -0.0467~ R_7 R_9 + 
0.0671~ R_9^2 \Big)\, .
\label{BRmu}
\eea
The SM values $~7.29\times 10^{-6}$, $4.89\times 10^{-6}$,
which correspond to
$\mathrm{BR}(B\to X_s\, e^+ e^-)$ and $\mathrm{BR}(B\to X_s\, \mu^+ \mu^-)$
respectively, are recovered by
setting $R_i=\tilde{R}_i=0$ in these formula. An important observation
from Eqs.(\ref{BRe}--\ref{BRmu}) is that the decay $b \to s\, l^+ l^-$ 
is quite sensitive to $R_{10}$ rather than the other variables. Therefore
any enhancement for $R_{10}$ could lead to significant changing in the 
prediction of BR of this decay without any consequences
on $b \to s \gamma$ decay, which mainly depends on $R_7$.   
It is worth noticing that the different sensitivity in $R_7$ in 
Eqs.(\ref{BRe}--\ref{BRmu}) is 
due to the fact that the coefficients proportional to $R_7$ 
come from integrating the $1/q^2$ pole 
(with $q^2$ the momentum square of the virtual photon)
of the magnetic operator $Q_7$. Therefore,
being the minimum value of $q^2$ proportional to the mass square 
of final leptons, the sensitivity to $R_7$ becomes larger in the 
electron channel.

We will also consider the lepton -- anti--lepton energy asymmetry (AS)
in the decay $b \to s\, l^+ l^-$ which is defined as
\be
\mathcal{A} = \frac{N(E_{l^-} > E_{l^+}) - N(E_{l^+} > E_{l^-})}
{N(E_{l^-} > E_{l^+}) + N(E_{l^+} > E_{l^-})},
\label{AS1}
\ee
where, for instance,
$N(E_{l^-} > E_{l^+})$ is the number of the lepton pairs whose 
negative charged member is more energetic in the $B$ meson rest frame
than its positive partner. As for the BRs we will consider 
the AS in Eq.(\ref{AS1}) 
integrated over non-resonant regions.
With the above parametrization we find \cite{misiak,sarid}
\bea 
\mathcal{A}_{ll}&=&\frac{0.48\times 10^{-6}}{\mathrm{R_{BR}
(B\to X_s\, l^+ l^-)} }\Big(1+0.911 R_{10} -0.00882~ R_{10}^2
-0.625~ R_7(R_{10}+1)\nonumber\\
&+&0.884~ R_{9}( R_{10}+1) \Big).
\label{AS}
\eea
where $R_{BR}=BR/BR^{SM}$.

Finally, regarding the $B \to X_s\, \gamma$ decay, we have used the following 
parametrization \cite{Emidio,sarid}
\bea 
{\rm BR}(B\to X_s\,\gamma) &=& (3.29\pm 0.33)\times 10^{-4} 
\left(1 + 0.622 R_7 + 0.090(R_7^2+\tilde{R}_7^2)
\right.\nonumber \\ 
&+& \left.0.066 R_8 + 0.019 (R_7 R_8 + \tilde{R}_7 \tilde{R}_8)
+ 0.002(R_8^2+\tilde{R}_8^2)\right)~,
\label{bsgPAR}
\eea 
where the overall factor corresponds to the SM value and its theoretical
uncertainty.


We start our analysis by revisiting the predictions for the rate of these
decays in the supersymmetric models with minimal flavor violation (such as 
the minimal SUGRA inspired model). 
In particular, we will show that the 
new bound on the Higgs mass~\cite{LEP} and the CLEO measurement for the 
BR of $B \to X_s\, \gamma$ decay ~\cite{CLEO}
\be
2.0 \times 10^{-4} < BRB \to X_s\, \gamma) < 4.5 \times 10^{-4}
\label{bsgEXP}
\ee
impose sever constraints on the parameter space of this class of models
and it is no longer possible to have deviations on the non-resonant BR of
$B \to X_s\, e^+ e^-$ and $B \to X_s\, \mu^+ \mu^-$
decays by more than $25\%$ and $10\%$ respectively, 
relative to their SM expectations. 

The main reason for that is the following.
As emphasized above, the main contributions to these processes are 
due to the operators $Q_7$, $Q_9$, and $Q_{10}$ where their Wilson 
coefficients are proportional to the mass insertions 
$(\delta^{u,d}_{23})_{LR}$ 
and $(\delta^{u,d}_{23})_{LL}$. However in the minimal SUGRA scenario, 
and due to the universality assumption upon the soft SUSY breaking 
parameters at GUT scale, the flavor transitions are suppressed by the 
smallness of the CKM angles and/or the smallness of the Yukawa couplings.
Moreover in this scenario, requiring the lightest Higgs mass to be 
$m_{h} > 110$ GeV implies that the universal gaugino masses $m_{1/2}$ 
has to be larger $250$ GeV. 
This leads to a heavy stop mass and hence a further suppression 
for the SUSY contribution to $B \to X_s\, l^+ l^-$ decays is found. 

\begin{figure}[tpb]
\dofigs{3.1in}{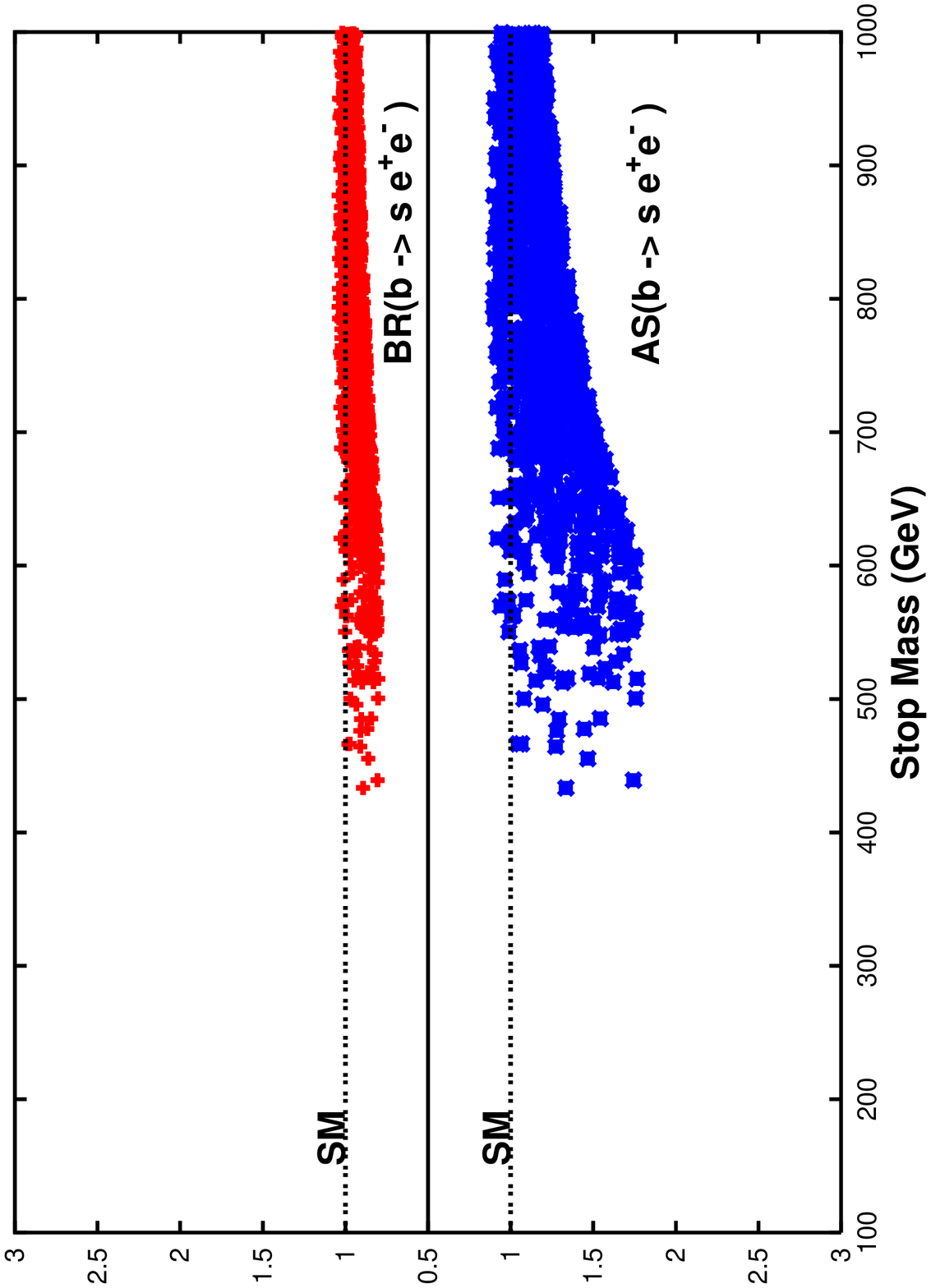}{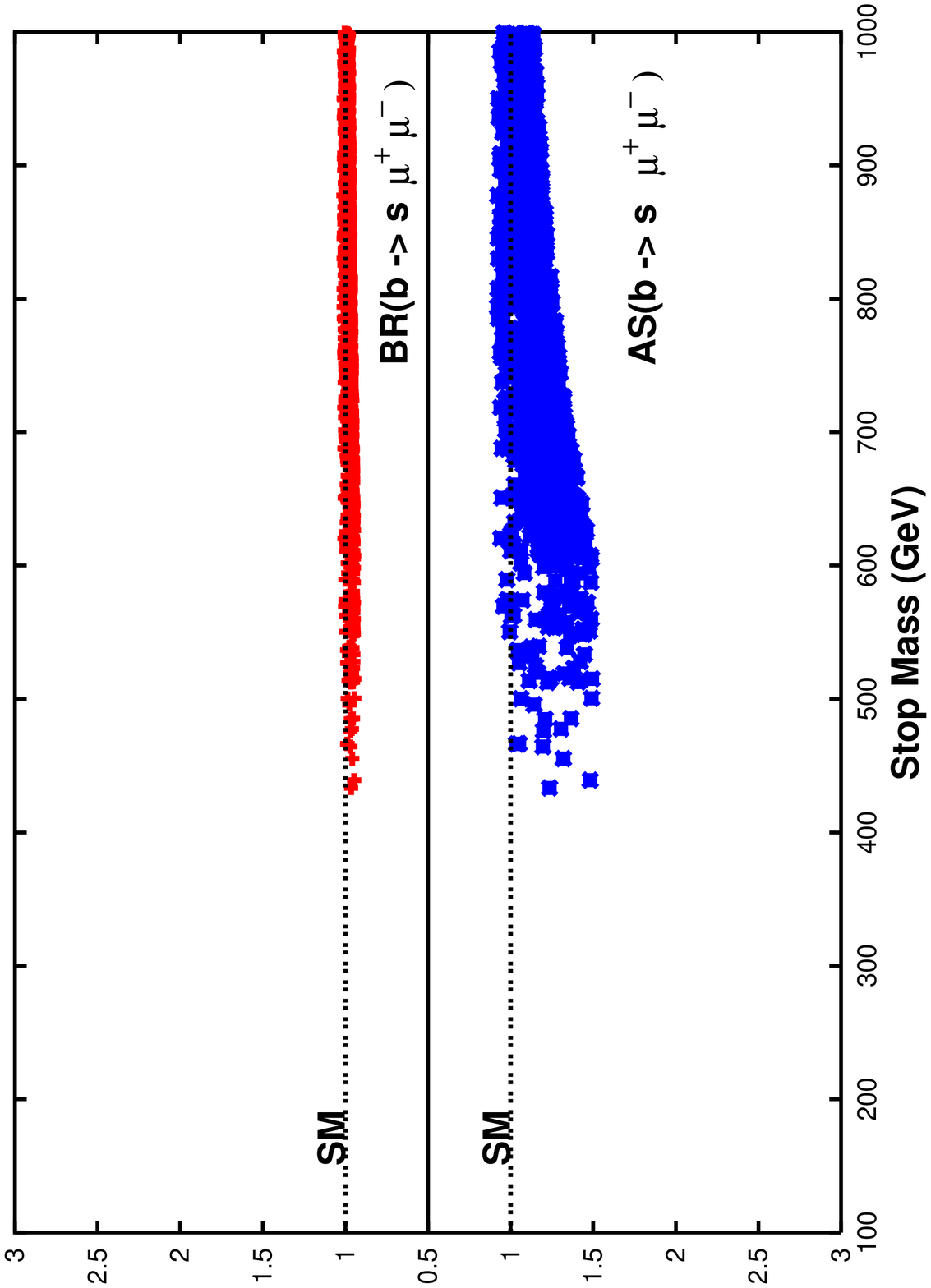}
\caption{{\small Branching ratio (BR) and energy asymmetry (AS) of 
$B \to X_s\, e^+ e^-$ and $B \to X_s\, \mu^+ \mu^-$ (normalized to the 
corresponding SM ones) versus the lightest stop mass in minimal SUGRA model.}}
\label{fig1} 
\end{figure}

In our analysis we present our results for a specific choice of 
the $sign(\mu)$. This choice corresponds to the one which gives
positive contributions to the $g-2$ of the muon, as it is
favoured by the new experimental results on $g-2$ \cite{g2EXP}.
Incidentally, this specific choice of $sign(\mu)$ is the one
for which the $B\to X_s \gamma$ constraints are less effective.
In Fig. \ref{fig1}, we present the scatter plots of the BR and 
the AS for the decay $B \to X_s\, e^+ e^-$ (which is the most 
sensitive semileptonic decay) and $B \to X_s\, \mu^+ \mu^-$ as a function of 
the lightest stop mass. 
In obtaining these figures, we varied the universal soft scalar 
mass $m_0$ and gaugino mass $m_{1/2}$ from 50 GeV up to 1 TeV.
The trilinear A-term is fixed to be $A_0=m_0$ and $\tan \beta$ vary in 
the range $[3, 40]$. 
In our numerical analysis we assume the radiative electroweak 
symmetry breaking and impose the current experimental bounds on the 
SUSY spectra. We have also imposed the constraints which come from 
requiring vacuum stability (necessary to ensure that 
the potential is bounded from below) and from avoiding charge and 
color breaking minima deeper than the real one.\footnote{
We stress that these last conditions may be automatically satisfied in
minimal SUGRA, while in generic SUSY models, like those we will consider below,
these conditions have to be explicitly checked. }
It turns out that the
present experimental limit on the lightest Higgs mass sets the most 
important constraint in minimal SUGRA models. 
In particular, it excludes the parameter 
space that leads to stop masses lower than 400 GeV.
It is clear that with such heavy stop masses the dominant contribution
to $b \to s\, l^+ l^-$, which comes from chargino exchanges,
is quite suppressed. 

\begin{figure}[tpb]
\dofigs{3.1in}{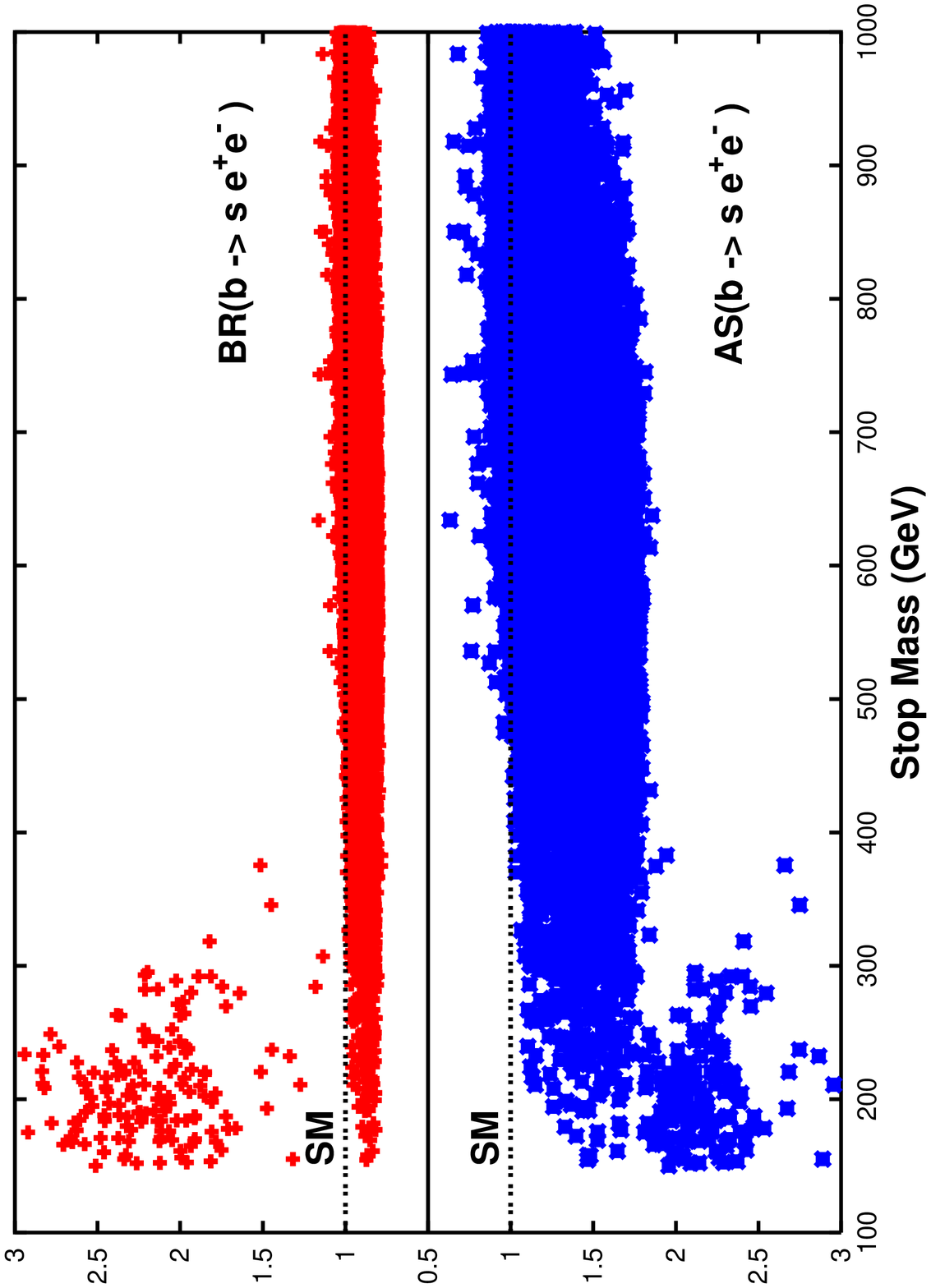}{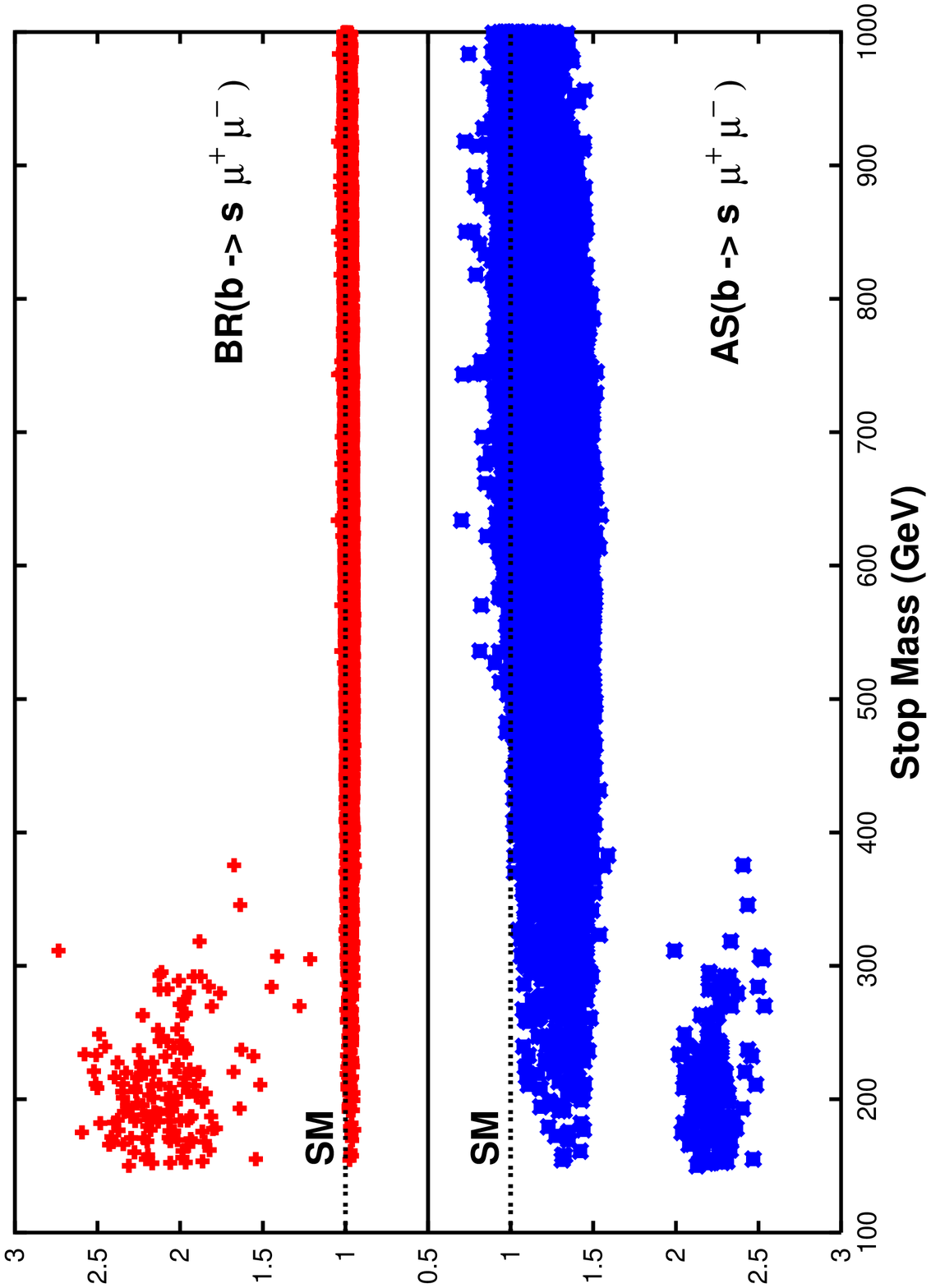}
\caption{{\small As in Fig. \ref{fig1}, but for SUSY model 
with non--universal soft breaking terms.}}
\label{fig2}
\end{figure}

\begin{figure}[tpb]
\dofigs{3.1in}{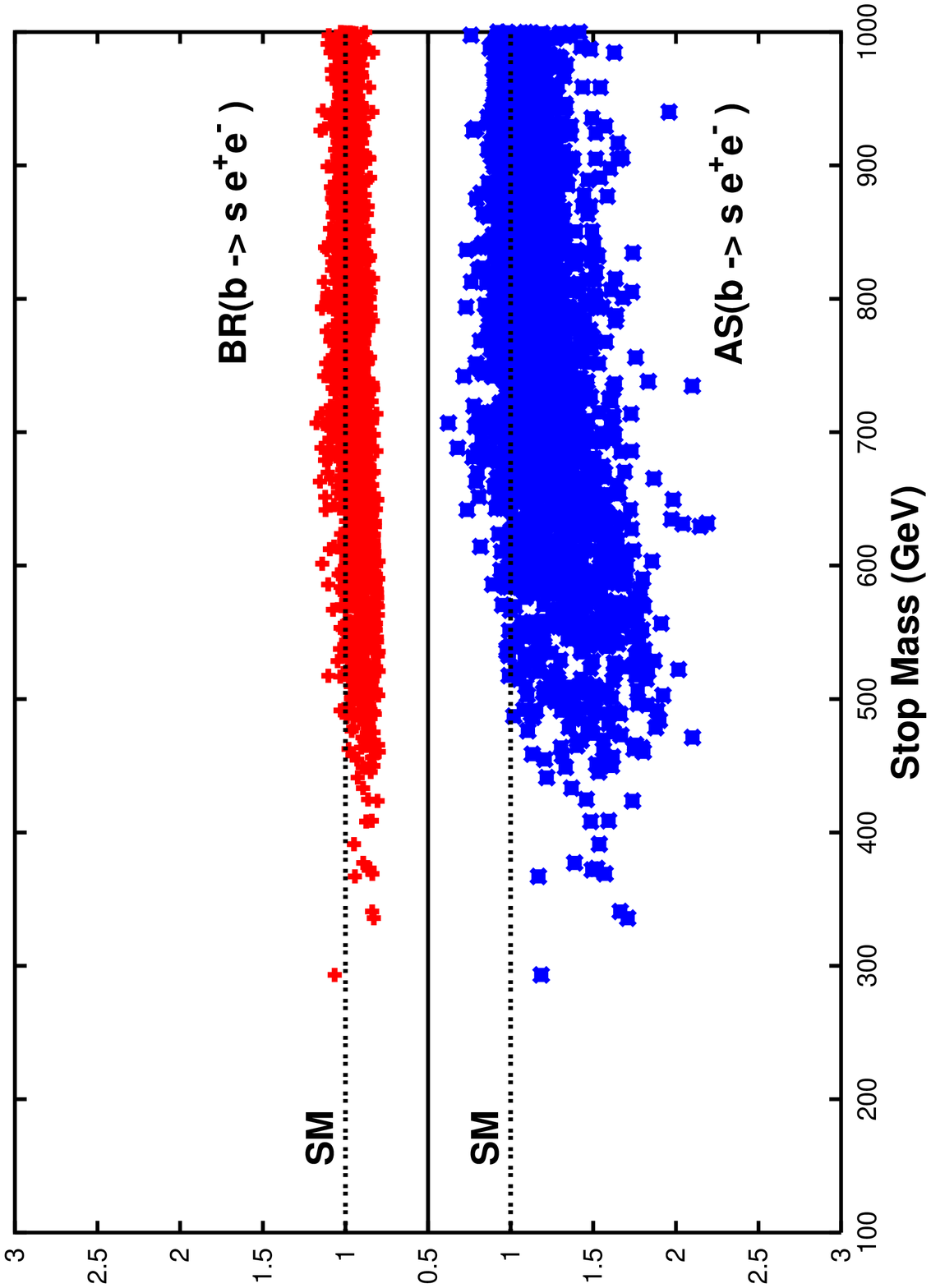}{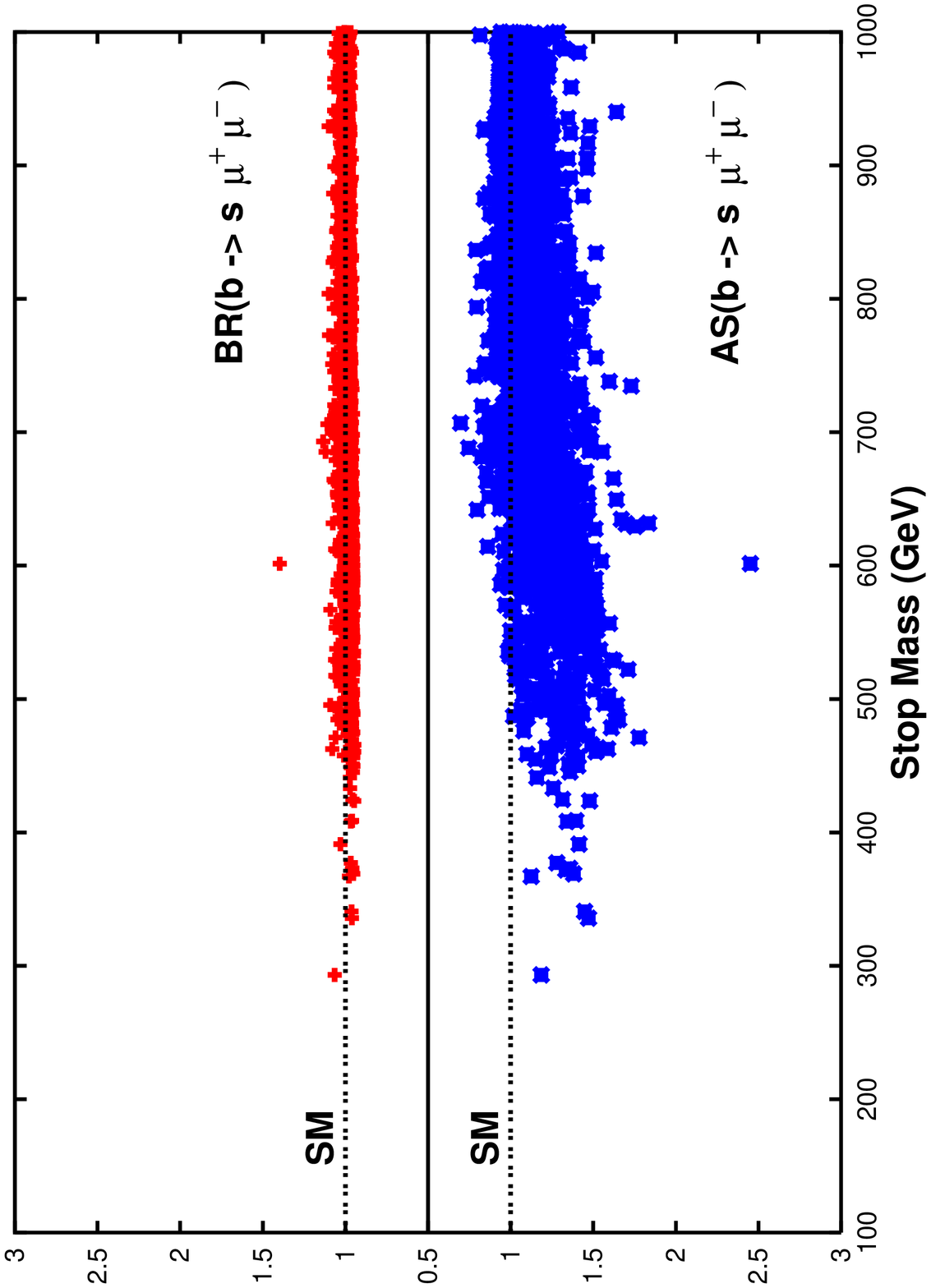}
\caption{{\small As in Fig. \ref{fig1}, but for a SUSY model 
with $U(2)$ flavour symmetry.}}
\label{fig3}
\end{figure}

As can be seen from Figs.\ref{fig1}--\ref{fig3}, the general trend of
this class of models, for this particular choice of $sign(\mu)$, 
is in a decreasing of BR and increasing of AS with respect to the SM 
expectations, in both universal and non-universal models.
The origin of this behaviour can be explained as follows.
As discussed above, the variations of BR and AS are mainly due to $R_7$.
For this choice of $sign(\mu)$ the
$B\to X_s \gamma $ constraints are less restrictive and 
mostly allow for negative values of $R_7$. 
Negative values of $R_7$ (in the range of [-1,0] ) will produce destructive
and constructive interferences in BR and AS respectively, 
as can be understood from the parametrizations in 
Eqs.(\ref{BRe})--(\ref{AS}).
However, we have also checked that for the other choice of $sign(\mu)$
the behaviour is opposite, giving an enhancement of BR and decreasing of 
AS, but with more moderate effects due to a stronger action of
$B\to X_s \gamma $ constraints.

In the large $\tan \beta$ region, where
chargino and Higgs contributions to $R_7$ are 
enhanced, ($R_9$ and $R_{10}$ are moderately
affected by $\tan \beta$), a sizeable changing in the BR
and AS of $B \to X_s\, l^+ l^-$ decays might arise.
Nevertheless, $R_7$ gives also the 
major contribution to the BR of 
$B \to X_s\, \gamma$, and by imposing the CLEO limits we dismiss such
large effects for $B \to X_s\, l^+ l^-$ decay.
Therefore, we can conclude that in SUSY models with 
universal soft breaking terms, it is not possible to get any significant 
enhancement for BR in $B \to X_s\, l^+ l^-$ decays, while 
a decreasing up to $25\%$ can be obtained in the electron channel. 
As explained above,
the decreasing of BR is reflected in a large enhancement of AS, in
particular 
up to $75\%$ and $50\%$ for the electron and muon channels respectively.
However, we will see that in general SUSY models,
mainly due to the non-universality in the scalar sector,
the Higgs bounds can be relaxed and larger deviations on BR and AS for 
$B \to X_s\, l^+ l^-$ decays can be achieved, deviations which correspond to
the allowed region of large negative values of $R_7$ (namely 
in the range of [-6,-4] ).


Now we turn to the most general supersymmetric extension of the SM. 
In particular, we will consider SUSY models with non--degenerate $A$--terms 
and non--universal soft scalar and gaugino masses.
Such models are naturally obtained from string inspired models 
~\cite{Brignole:1997fb}
and some aspect of their phenomenological implications have been recently 
studied. 
Note that the squark mass matrices are often diagonal in
string inspired models and this is what we will adopt here. 
Generic SUSY models might also have
non--universality in the off-diagonal terms of squark mass matrices. 
Nevertheless, these off-diagonal terms 
are severely constrained 
by $\Delta M_K$, $\Delta M_B$, and $\varepsilon_K$.
Models with flavor symmetries naturally avoid such constraints. 
We will consider later a model with $U(2)$ flavor symmetry as an 
example for this class models.
 
In order to parametrize the non-universality of
a large class of string inspired models 
(with diagonal soft-breaking terms in the sfermion sector),
we assume here the following soft scalar masses, gaugino masses $M_a$
and trilinear couplings:
\bea
M_a &=&\delta_a~ m_{1/2},~~ a=1,2,3,\\
m_Q^2 &=& m_L^2 = m_0^2~ diag\{1,1,\delta_4\},\\
m_U^2 &=& m_0^2~ diag\{1,1,\delta_5\},\\
m_D^2 &=& m_E^2 = m_0^2~ diag\{1,1,\delta_6\},\\
m_{H_1}^2 &=&m_0^2~ \delta_7,\, \, m_{H_2}^2 =m_0^2~ \delta_8,\\
A^u&=& A^d = A^l = m_0 \left(
                    \begin{array}{ccc}
                        a_{11} & a_{12} & a_{13}\\
                        a_{21} & a_{22} & a_{23} \\
                        a_{31} & a_{32} & a_{33}
                        \end{array} 
                        \right).        
\eea
where the parameters $\delta_i$ and $a_{ij}$ can vary
in the $[0,1]$ and $[-3,3]$ ranges respectively.
It has recently been emphasized that these models could be free from the 
EDMs constraints and also have testable implications 
for the CP violation experiments~\cite{nonuniversalA}. 
In Ref.\cite{Emidio}, the prediction for the BR of 
$B \to X_s\, \gamma$ decay has been considered in two 
representative examples for this class of models and it was found that 
$B \to X_s\, \gamma$ does not essentially constrain 
the non--universality of $A$--terms.

In our convention for the trilinear couplings, the $A$ terms are defined
such that $\hat{A}_{ij}= A_{ij} Y_{ij} $ (indices not summed) and $Y_{ij}$
are the corresponding Yukawa couplings. We assume that the Yukawa matrices
at EW scale are given by
\be
Y^d = \frac{1}{v_1} V^*_{CKM} \mathrm{diag}(m_d, m_s,m_b), ~~~
Y^u = \frac{1}{v_2} \mathrm{diag}(m_d, m_s,m_b) V^T_{CKM}\;,
\ee
For any value of the parameters
$m_0$, $\delta_i$, $m_{1/2}$, $a_{ij}$ at GUT scale, 
and $\tan \beta$ 
(we determine the $\mu$ and $B$ parameters from the electroweak 
breaking conditions)
we compute the relevant SUSY spectrum and 
interaction vertices at low energy needed for the calculation of the
$b \to s\, l^+ l^-$ decay amplitudes. In order to connect
the high energy SUSY parameters, gauge and Yukawa couplings
with the corresponding low energy ones,
we have used the most general renormalization group 
equations in MSSM at 1-loop level. As stated above, we impose the 
current experimental bounds on SUSY spectrum, in particular
lightest Higgs mass $m_h > 110$ GeV, and $B\to X_s\gamma$
constraints in Eq.(\ref{bsgEXP}).

In Fig. 2 we present scatter plots for the BR and AS
for the $B \to X_s\, e^+ e^-$ and $B \to X_s\, \mu^+ \mu^-$ 
decays versus the lightest stop mass.  
As for the universal models, we varied the fundamental mass parameters 
$m_0$,~ $m_{1/2}$ from 50 GeV up 1 TeV,
and $\tan \beta$ in the range $[3,40]$. The parameters
$\delta_i$ and $a_{ij}$ have been also randomly
selected in the ranges $[0,1]$ and  $[-3,3]$ respectively. 
It is worth mentioning that the gluino contributions are 
negligible in the universal limit and the non--universality in the $A$--terms 
is essential for enhancing such contributions. Moreover, with 
non--universality in the gaugino masses we can have light chargino 
and stop masses close to their
experimental limit and the Higgs mass bound satisfied. In this 
region of parameter space indeed
the chargino contributions to $R_7$, $R_9$ and $R_{10}$ are enhanced.
However, it is noted that in all the parameter space, 
$R_{9,10}$ are much smaller than $R_7$ and 
still the main contributions to these processes are due to $R_7$ which 
also gives the main contribution to the $B \to X_s\, \gamma$ decay. 

As can be seen from Fig. \ref{fig2} there is a disconnected region 
of points, for stop masses lighter than 300 GeV, 
where very large enhancements in both BR and AS are reached.
In particular, a factor 3 and 2.5 of enhancements in both 
BR and AS are obtained for electron and muon channels respectively.
This region corresponds to 
the large (negative) SUSY contributions to
$R_7$, roughly in the range of [-6,-4], obtained for $\tan\beta > 30$. 
Nevertheless, these huge enhancements 
belong to the less populated areas of scatter plots which means that
a larger amount of fine tuning between the SUSY parameters is needed
in this case.

The more populated areas in Fig.\ref{fig2} correspond to the
other (disconnected) range of allowed values for $R_7$, namely $-1< R_7<1 $.
In this region, the $B \to X_s\, \gamma$ constraints reduce the 
enhancements (with respect to the SM one) on the BR
of $B \to X_s\, e^+ e^-$ to be less than $20\%$ and the decreasings up to 
$25\%$. More moderate effects are obtained 
for the muon channel, since it is less sensitive to $R_7$.
In correspondence to these variations on the BR, larger
effects are obtained for the AS. 
In particular up to $75\%$ and $50\%$ enhancements
in the AS for electron and muon channel respectively, 
while a more moderate increasing
(about $40\%$ and $25\%$ respectively) are expected.

Now we compare our results with the model independent 
analysis of Ref.~\cite{masiero}, based on a low energy approach.
Using the mass insertion approximations and general MSSM at low energy 
it was shown in Ref.~\cite{masiero} that the SUSY contributions 
to BR and AS of the semileptonic decays can get 
maximum enhancement (up to $4\times 10^{-5}$ for the $BR(B \to X_s\, e^+ e^-)$ 
i.e 4 times the SM value). However, this needs the following values for 
the mass insertions $(\delta_{23})$:
\be
(\delta^{u,d}_{23})_{LL} \simeq - 0.5, \, ~~ \,  
(\delta^{u}_{23})_{LR} \simeq 0.9.
\ee
Such values can be obtained only in a very small region of the parameter 
space of the SUSY models with non--universal soft terms,
specially after imposing the electroweak breaking conditions, the 
new bounds on the Higgs mass, and $B\to X_s\, \gamma$ contraints.
However, we found that in general the typical values of these mass 
insertions are  $\vert(\delta^{u,d}_{23})_{LL} \vert \simeq 10^{-2}, \,  
(\delta^{u}_{23})_{LR} \simeq 10^{-3}$. This, indeed, leads to a BR
for the $B \to X_s\, e^+ e^-$ decay of order $10^{-6}$ with at most $20\%$ 
enhancement than the SM value.

 
Finally we proceed to consider SUSY models with non--abelian flavour symmetry. 
This class of models has a flavour structure in the soft scalar masses, 
and hence, the $LL$ sector contains larger mixing than what 
is found in the previous models with diagonal squark masses. 
As mentioned, the $\Delta M_K$ and $\varepsilon_K$
impose sever constraints on the squark mixing, namely $\sqrt{\vert
Re(\delta_{12})^2_{LL} \vert} \lsim 10^{-2}$ and 
$\sqrt{\vert Im(\delta_{12})^2_{LL} \vert} \lsim 10^{-3}$ respectively
\cite{FCNCsusy}.

Here as an illustrative example, we consider a model
based on a $U(2)$ symmetry acting on the two light 
families~\cite{silvestrini} where 
the above mentioned constraints are satisfied.
In this case, the Yukawa textures, at GUT scale, 
are given by~\cite{silvestrini}
\begin{equation}
Y_u = \frac{m_t}{v \sin \beta} \left(
\begin{array}{ccc}
0 & c \varepsilon \varepsilon' & 0 \\
- c \varepsilon \varepsilon' & 0 & a \varepsilon \\
1 & b \varepsilon & 1
\end{array} \right) ;\;
Y_d = \frac{m_b}{v \cos \beta} \left(
\begin{array}{ccc}
0 & \frac{\varepsilon'}{\sqrt{1+ \rho^2 k^2}} & 0 \\
- \varepsilon' & 0 & a \varepsilon \\
1 & \rho & 1
\end{array} \right) ;\;
\end{equation}
and the squark mass matrices take the form
\begin{eqnarray}
M_Q^2 &=& m_{3/2}\left(
\begin{array}{ccc}
1 & 0 & \alpha \varepsilon \varepsilon' \\
0 & 1 & 0 \\
\alpha^* \varepsilon \varepsilon' & 0 & r_3
\end{array} \right) ;\;
M_D^2 = m_{3/2} \left(
\begin{array}{ccc}
1 & 0 & \alpha' \varepsilon \varepsilon' \\
0 & 1+ \lambda \vert \rho \vert^2 & \beta \rho^* \\
\alpha'^* \varepsilon \varepsilon' & \beta^* \rho & r'_3
\end{array} \right) ;\;\nonumber\\
M_U^2 &=& m_{3/2} \left(
\begin{array}{ccc}
1 & 0 & \alpha'' \varepsilon \varepsilon' \\
0 & 1 & 0 \\
\alpha''^* \varepsilon \varepsilon' & 0  & r''_3
\end{array} \right) .\;
\end{eqnarray}
The 
definition of the parameters appearing in these matrices can be found in 
Ref.\cite{silvestrini}. The important feature of the flavor structure of this
model is the presence of a large mixing between the second and the 
third generation which would have significant effect on enhancing 
the BR and AS of $b\to s\, l^+ l^-$ decays. 
However, this mixing essentially enhances $R_7$ 
which means enhancing for the BR of $B \to X_s \gamma$ decay 
as well. Therefore imposing the $B \to X_s \gamma$ constraints this leads to a 
similar prediction to that we obtained with the previous model.
 
In Fig.3 we display the predictions of this model for BR 
and AS of $B \to X_s\, e^+ e^-$ and $ B \to X_s\, \mu^+ \mu^-$ 
decays versus the lightest stop mass.
As in the previous models we have considered, most of the parameter 
space (favored by the $B \to X_s \gamma$ and other constraints) leads to
decreasing in the BR and increasing of AS.
This is due to the fact that even for these models the major effect in the
variation is due to $R_7$.
The large enhancements in BR and AS, obtained in the other scenario with
large and negative contributions to $R_7$, are not very likely to
show up. This is mainly due to the constraints on the Higgs mass,
which prevent stop masses to be lighter than 300 GeV.

\section*{\bf \normalsize Conclusions}

We have analyzed the predictions for the 
inclusive semileptonic decays $B \to X_s\, l^+ l^-$ 
in different SUSY models. In particular, we have considered SUSY models with 
minimal flavor violation, non--degenerate $A$--terms and non--universal
soft scalar and gaugino masses, and finally SUSY models with non--abelian 
symmetry that leads to a flavor structure for the soft scalar masses.
We showed that in all these models the major effect on the variations
of $B \to X_s\, l^+ l^-$  decays, with respect to their SM expectations,
is due to the SUSY contributions to the magnetic dipole operator 
parametrized by $R_7$ (which also give the major contribution to 
the inclusive $B \to X_s\, \gamma$ decay). The SUSY contributions 
to the semileptonic operators is almost negligible. 

We found that the general trend of our results, favoured by the CLEO 
$B\to X_s \gamma$ constraints and Higgs mass bound, 
is in decreasing the non-resonant BR and increasing 
the AS. Nevertheless, only non--universal models 
can have chances to get very large enhancements in BR and AS.
In particular, in this case up to 3 and 2.5 time enhancements of BR and 
AS with respect to the SM expectations can be obtained in the electron 
and muon channel respectively.

%
\section*{\bf \normalsize Acknowledgements}
We would like to thank A. De Andrea, G. Isidori, and A. Polosa
for useful discussions.
E.G. acknowledges the kind hospitality of the CERN Theory Division
where part of this work has been done.
The work of S.K. was supported by PPARC.

%

\end{document}